\newcommand{\rem}[1]{}
\documentclass[prb,twocolumn,showpacs,preprintnumbers,amsmath,amssymb,floatfix]{revtex4}
\usepackage{hyperref}
\usepackage{hypernat} 
\usepackage[dvips]{graphicx}

\begin{document}

\title{Comment to ``Imaging the atomic orbitals of carbon atomic chains
  with field-emission electron microscopy''}

\author{Nicola Manini and Giovanni Onida}
\affiliation{
ETSF and Dipartimento di Fisica, Universit\`a degli Studi di Milano,
Via Celoria 16, 20133 Milano, Italy
}

\date{Dec 22, 2009}
                                     
\begin{abstract}
The observation of a stable doublet pattern in the field-emission electron
microscopy of a linear atomic chain requires a stable mechanism breaking
the axial symmetry, which is not identified correctly by Mikhailovskij {\it
  et al.}  [Phys.\ Rev. B {\bf 80}, 165404 (2009)].
Using microscopic calculations, we attribute the observed pattern to the
symmetry breaking produced by the ligand where the chain is attached, plus
carbon $\pi$-bonding alternation.
\end{abstract}


\maketitle

Very recent field-emission electron microscopy (FEEM) experiments
\cite{Mikhailovskij09} show compelling evidence of emission from individual
molecular orbitals (MO) of free-standing carbon atomic chains.
The data are consistent with alternatively a single 
spot or a
doublet of spots, which the authors identify as emission from $s$-type or
$p$-type MO respectively.
$s$ / $p$ labeling refers to the angular momentum of these electronic
states around the chain axis, usually denoted by $\sigma$ / $\pi$ in the
quantum-chemistry literature.
An electronic wave function of the axially symmetric chain depends on the
azimuthal angle $\varphi$ around the chain axis as $\exp(\pm i n\varphi)$,
with $n=0$, 1 for $s$, $p$ states respectively, see Eq.~(5) of
Ref.~\onlinecite{Mikhailovskij09}.

Contrary of the arguments of Ref.~\onlinecite{Mikhailovskij09}, $n=1$
does {\em not} imply the existence of angular nodes.
If $\exp(\pm i \varphi)$ was indeed the angular dependency, FEEM would
display an axially symmetric (possibly ring-shaped) pattern.
Of course, a perturbation could localize the electron in the azimuthal
angle, by inducing a splitting between the $\cos(\varphi-\varphi_0)$ and
$\sin(\varphi-\varphi_0)$ components of the $p$ state, which do in turn
display nodes.
However, any sort of fluctuating perturbation, such as those associated to
the thermal oscillation of the free standing 
chain, would produce a rapidly
fluctuating
phase $\varphi_0$, with the eventual result of
an axially symmetric
pattern.
A stable doublet could arise in FEEM only due to a symmetry-breaking
perturbation that {\em locks the node at a fixed $\varphi_0$}.
This perturbation needs to be sufficiently large and stable to overcome the
tendency of thermal vibrations and quantum kinetic energy to wash out all
localization effects.

What can induce such a symmetry-breaking perturbation in a
several-atom-long free-standing chain?

\begin{figure}
\centerline{
{\small (a)}\includegraphics[height=37mm,angle=0,clip=]{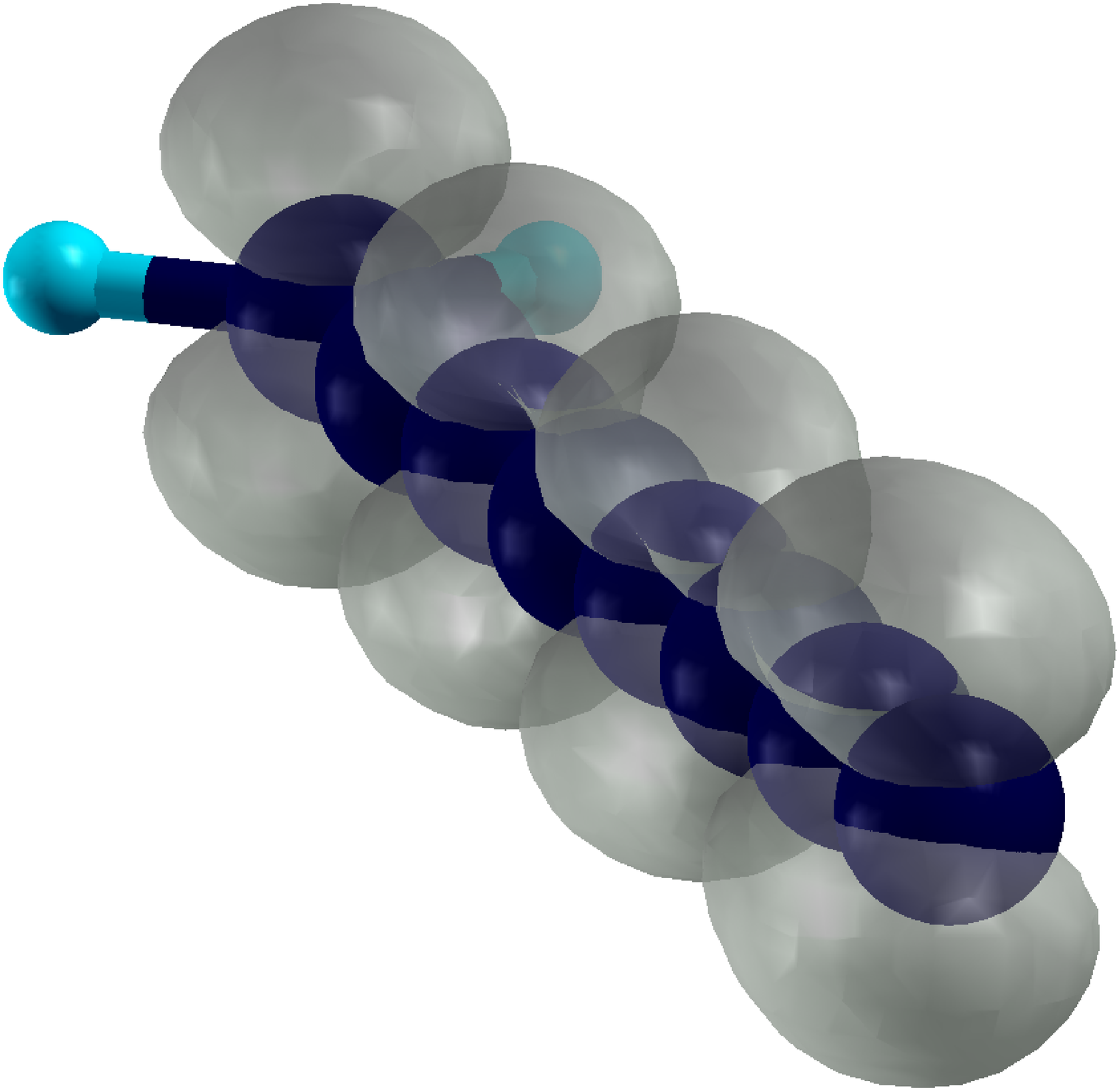}
\hfill
{\small (b)}\includegraphics[height=37mm,angle=0,clip=]{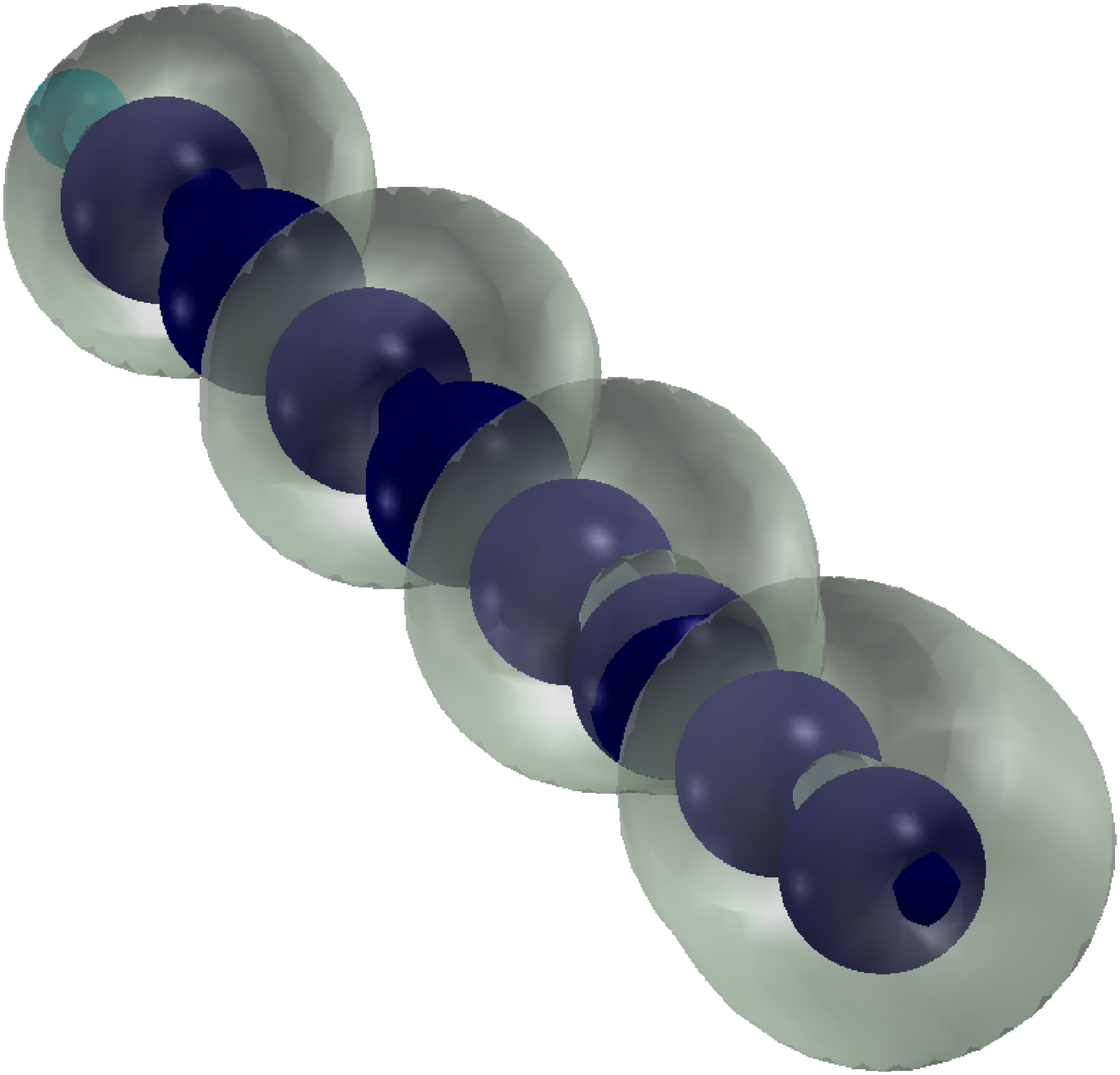}
}
\caption{\label{fig} (Color online)
The HOMO (grey transparent level surface surrounds the region where
$\left|\psi_{\rm HOMO}\right|^2>0.004\;a_0^{-3}$) of a linear chain of
carbon atoms (dark balls) attached to (a) 2 ligands or (b) one ligand
(hydrogen, clear balls).
}
\end{figure}

The answer
is: the anchored end of the chain.
Whenever this end binds an $sp^2$ hybridized atom of the carbon tip,
the chain acquires a partial cumulenic character\cite{Ravagnan09}.
All bonds are double, and the memory of the orientation of the $sp^2$
termination propagates along the chain through an alternating orientation
of $\pi$ bonds \cite{Gu08}.
We illustrate this point by carrying out standard calculations within the
Density-Functional theory in the local-density approximation (DFT-LDA) with
a plane-waves package \cite{espresso2009,espresso-avail} using default
ultrasoft pseudopotentials, and wavefunction/charge cutoffs of 15/120
Hartree.
We relax the atomic positions until the largest residual force is
less than $10^{-4}$~Hartree/$a_0$ (8~pN).
For a 8-atom chain, Fig.~\ref{fig}a shows the highest-occupied molecular
orbital (HOMO), made of the four $\pi$ bonds oriented normally to the
$sp^2$ plane.\footnote{
The remaining three $\pi$ bonds point horizontally, in the $sp^2$ plane.
}
As a result, emission from the terminal atom is dominated by this
asymmetric $\pi$ orbital, and displays a clear nodal plane which is
parallel/antiparallel to the $sp^2$ plane, depending on the chain being
composed by an even/odd number of atoms.

Alternatively, the anchored end, rather than $sp^2$, could have $sp^3$
hybridization (e.g.\ a generic atom in the ``bulk'' of a graphene fragment
or fullerene or nanotube cap, represented by a simple H atom in
Fig.\ref{fig}b).
Such a ligand would induce a polyyinic-type electronic structure of the
chain.
As a result, a highly dimerized configuration, with alternating
single/triple bonds and an essentially unbroken axial symmetry is realized.
However, even in this case, eventually the HOMO is a $\pi$ bonding orbital.
The main difference is that as the cylindrical symmetry is unbroken, a
cylindrically symmetric ``doughnut-shaped'' orbital is generated, precisely
of the type described by Eq.~(5) of Ref.~\onlinecite{Mikhailovskij09}.
Such a state vanishes {\em along the molecular axis}, as can be seen in
Fig.\ref{fig}b, but, due to thermal vibrations of the chain, it is unlikely
that its FEEM pattern could distinguish it from that of a singlet
$\sigma$-type orbital.\footnote{
For increasing chain length, the dangling-bond terminal $\sigma$ orbital
remains energetically below the HOMO level, with an increasing separation
from the HOMO, in the $100-400$~meV range.
}
Indeed higher currents in excess of 100~pA can induce switching from one
type of FEEM pattern to the other, most likely by exciting a jump of the
chain attaching point, not unlike those observed in recent experiments
\cite{Jin09,Chuvilin09} under the beam of an electron microscope.



\end{document}